\documentclass[superscriptaddress,twocolumn,showpacs,showkeys,amsmath,amssymb,nofootinbib]{revtex4}
\usepackage{graphicx}
\usepackage{dcolumn}
\usepackage{bm}

\begin{document}

\title{Quantum Monte-Carlo method applied to Non-Markovian barrier transmission}

\author{Guillaume Hupin}
\affiliation{GANIL, Bd Henri Becquerel, BP 55027, 14076 Caen Cedex 5, France}

\author{Denis Lacroix}
\affiliation{GANIL, Bd Henri Becquerel, BP 55027, 14076 Caen Cedex 5, France}


\begin{abstract}
In nuclear fusion and fission, fluctuation and dissipation 
arise due to the coupling of collective degrees of freedom with internal excitations.
Close to the barrier, 
both quantum, statistical and non-Markovian effects are expected to be important. In this work, a new approach 
based on quantum Monte-Carlo addressing this problem is presented.
The exact dynamics of a system coupled to an environment is replaced by a set of stochastic
evolutions of the system density. The quantum Monte-Carlo method is applied to  systems 
with quadratic potentials. In all range of temperature and coupling, the stochastic method 
matches the exact evolution showing that non-Markovian effects can be simulated 
accurately. A comparison with other theories like Nakajima-Zwanzig or Time-ConvolutionLess 
ones shows that only the latter can be competitive if the expansion in terms of
coupling constant is made at least to fourth order. 
A systematic study of the inverted parabola case is made at different 
temperatures and coupling constants. The asymptotic passing probability is estimated in different approaches 
including the Markovian limit.
Large differences with the exact result are seen in the latter case or when only second order 
in the coupling strength is considered as it is generally assumed in nuclear transport models. 
On opposite, if fourth order in the coupling or quantum 
Monte-Carlo method is used, a perfect agreement is obtained. 
\end{abstract}
\maketitle

\section{Introduction}

To understand nuclear reactions, the dynamics of nuclei is often replaced 
by few selected collective degrees of freedom expected to contain important information on  
the dynamic. This is for instance the case in fusion reactions where the relative distance and/or mass asymmetry 
is retained \cite{Sch84,Fro96}. Another example is provided by the fission process which is often 
treated as a trajectory in an energy landscape function on different deformation parameters \cite{Ber84,Mol70}. Although the 
evolution is projected onto few variables, other internal degrees of freedom may play 
an important role to understand 
the onset of dissipation or fluctuation phenomena \cite{Abe96}.  To treat these effects, 
the relevant degrees of freedom should be regarded as an Open Quantum System coupled to an environment which simulates the internal dynamics.

To include dissipation in collective space, two important simplifications are often made. First, most of 
current models treating fusion/fission neglect quantum effects and consider a classical treatment \cite{Ari99,Abe00,Abe02,She02}. Such an approximation is however expected to be valid only if the internal 
excitation is high and therefore is not expected to hold close to or below the Coulomb barrier. As it is discussed in ref. \cite{Dia08}, 
a proper treatment of quantum and decoherence effects might be crucial in this region.   
Second, 
when the time scale associated to collective dynamics cannot be dissociated from the one of the
environment, "non-Markovian" (also called "memory") effect should also be properly treated \cite{Boi06,Mat00}. 
Large effort is now devoted to account for both quantal and non-Markovian effects in nuclear reactions 
\cite{Rum03,Tak04,Ayi05,Sar07,Was07,Sar08} and more generally in open quantum systems \cite{Bre02}.

Recently, the description of open quantum systems by stochastic methods has received much attention \cite{Ple98,Bre02,Sto02}.
In the Markovian limit, several methods have been proposed to treat fluctuation and dissipation 
starting from a quantum master equation on the system density \cite{Dal92,Dum92,Gis92,Car93,Cas96,Rig96,Ple98,Gar00,Bre02}.  
These methods have been extended also to  treat non-Markovian effects like in Quantum State 
Diffusion (QSD) \cite{Dio96,Dio98,Str99,Str05} or quantum Monte-Carlo (QMC) methods \cite{Pii08}.
Several groups have shown that these effects could eventually be simulated using Feynman-Vernon 
influence functional \cite{Sto02,Sha04} or directly stochastic master equations  
\cite{Lac05,Lac08b}.

In this work, we apply the stochastic formulation proposed in ref. \cite{Lac08b} to the case of quadratic potentials coupled 
to a heat-bath, the so-called Caldeira-Leggett model \cite{Cal82}. The case of inverted potential is the first step towards 
realistic 
situations like fusion or fission. The aim of the present work is threefold. First, to introduce the new QMC method and apply it 
for potentials with barriers similar to those appearing in fusion/fission processes. Second, we show that the exact quantum 
Monte-Carlo method can be rather accurate to treat the dissipative dynamics of a quantum system. Last, 
we also present a comparison of this theory
with other methods based on projection, namely Nakajima-Zwangig (NZ) and Time-ConvolutionLess (TCL) \cite{Nak58,Zwa60,Has77,Shi77}, 
which are actually widely used to treat non-Markovian effects. Doing so, we show that only TCL up to at least fourth orders in 
the coupling constant can provide a competitive theory. 
The paper is organized as follows. In section 
\ref{sec:exact}, the ingredients and properties of the quantum Monte-Carlo 
approach are discussed and the link with functional integral
is precised. In section \ref{sec:appli}, the method is first illustrated to the case of parabolic 
potential. Then, the passing probabilities are estimated for the inverted 
parabola case.

\section{Quantum Monte-Carlo method}
\label{sec:exact}
 
Our starting point is a system (S) interacting with a surrounding 
environment (E). We assume here that the total  system (S+E) is described by the Hamiltonian 
\begin{eqnarray}
H = H_S  + H_E + V. \label{eq:hamil}
\end{eqnarray} 
$H_S$ (resp. $H_E$) acts on the system (resp. env.) only while $V$ induces a coupling between the two sub-systems. 
Starting from an initial total density $D(0)$, the dynamical evolution is given by the Liouville von-Neumann equation on the density:
\begin{eqnarray}
i\hbar \frac{d D(t)} {dt} &=& [H,D(t)].
\label{eq:evoldtot}
\end{eqnarray}
In many physical situations, the total number of degrees of freedom to follow in time prevents from solving exactly this equation. 
One of the leitmotiv of Open Quantum System (OQS) theory is to find accurate approximations for the system evolution without 
following explicitly irrelevant degrees of freedom associated to the environment and therefore reduces the complexity of the 
initial problem. 
Conventional strategy to treat dissipation and fluctuation in an open quantum system 
is to reduce the information to the 
system density only $\rho_S(t) = {\rm Tr_E}(D(t))$ while accounting approximately for environment effect. Here, we use a different strategy, the 
dynamics of the total system is first replaced by a set of stochastic evolutions where the total density remains separable 
along each path, 
i.e. $D = \rho_S(t)\otimes \rho_E(t)$. Then, the stochastic evolution of the environment is projected onto the relevant degrees 
of freedom to obtain a closed equation for the system density. It is shown that the new approach provides a proper treatment of dissipation and fluctuation for a system coupled to a surrounding heat-bath.  


\subsection{Quantum Monte-Carlo formulation of Open Quantum Systems}
\label{sec:flex}

Recently, new stochastic formulations \cite{Sha04,Bre04c,Lac05,Lac08} have been developed to 
study the system+environment problem that avoid evaluation of non-local memory 
kernels although non-Markovian effects are accounted for exactly 
(see also \cite{Bre04a,Bre04b,Bre04c,Lac05,Zho05,Bre07,Lac08b}). One example of such 
a theory based on quantum Monte-Carlo method is presented here.

Hereafter, it is assumed that the coupling  
is separable: $V = Q \otimes B$ where $Q$ and $B$ act on the system and environment 
respectively. For simplicity, initial separable density is considered, i.e.
$D(0) = \rho_S(0) \otimes \rho_E (0)$. 
{We want to replace the evolution 
of the total density (Eq. (\ref{eq:evoldtot})) by an ensemble of stochastic evolutions 
of both the system and environment
such that:
\begin{eqnarray}
\left\{
\begin{array} {lll}
d\rho_S &=& \frac{dt}{i\hbar}[H_S, \rho_S]  +  
d\xi_S Q  \rho_S + d\lambda_S 
\rho_S Q
\\
\\
d\rho_E &=& \frac{dt}{i\hbar}[H_E ,\rho_E]  +  
d\xi_E {B} \rho_E 
+ d\lambda_E \rho_E {B}  
\end{array}
\right.
\label{eq:stocmfsimple}
\end{eqnarray}
where $d\xi_{S/E}$ and $d\lambda_{S/E}$ denote Markovian Gaussian stochastic variables with zero means and where we use 
Ito convention of stochastic calculus \cite{Gar85}. In the following, we assume in addition that 
\begin{eqnarray}
\overline{d\xi_S d\lambda_E} &=& \overline{d\lambda_S d\xi_E} = 0, \label{eq:noise2}
\end{eqnarray}    
where the overline denotes the stochastic average. Along each path, the total density remains separable, i.e. $D(t) = \rho_S(t) \otimes \rho_E (t)$. Starting from 
such a density, at time $t+dt$, the average evolution deduced from Eq. (\ref{eq:stocmfsimple}) is given by 
\begin{eqnarray}
\overline{ d D(t)} &=& \frac{dt}{i\hbar} [h_S + h_E , D(t)] \nonumber \\
&& + \overline{d\xi_S d\xi_E} ~(Q\otimes B)  D(t) + \overline{d\lambda_S d\lambda_E} ~D(t)(Q\otimes B). \nonumber
\end{eqnarray}  
Therefore, under the condition
\begin{eqnarray}
\overline{d\xi_S d\xi_E} &=& \frac{dt}{i\hbar},~~~~\overline{d\lambda_S d\lambda_E} = -\frac{dt}{i\hbar}, \label{eq:noise1} 
\end{eqnarray}  
the average evolution over the separable densities that evolve according to Eq. (\ref{eq:stocmfsimple})
identify with the exact Liouville von Neumann equation of motion (Eq. (\ref{eq:evoldtot})). 
{The possibility to use simple Gaussian noises to incorporate the environment effect might appear 
surprising. Indeed, noises used in standard approaches for Open Quantum Systems generally reflect 
properties of the environment. It should be however kept in mind that such environment dependent 
noises appear once the environment dynamics has been projected out on the system density evolution.
Anticipating the discussion of section \ref{sec:master}, once such a projection has been made, 
the Gaussian noises introduced here, transform into new random variables that explicitly depend on the environment properties.
}

The discussion above for one 
time step can then be iterated 
to show that the exact dynamics of a system+environment 
could be replaced by an average over an ensemble of stochastic evolutions of separable densities \cite{Sha04,Lac05,Lac08b}. To be really useful, mainly two 
difficulties should be overcome (i) in general, the environment corresponds to a large number of degrees of freedom that could not be followed explicitly in time. (ii) the numerical implementation of such a theory is possible only if the statistical errors do not growth too fast during 
the time-evolution. This statistical errors are directly connected to the number of trajectories necessary to accurately describe the physical 
process. Difficulty (i) is solved in the next section by projecting out the effect of the environment on the system leading to a closed 
equation for the system density only. Let us first concentrate on statistical errors.
At any time, a measure of the statistical fluctuation around the average trajectory is given by 
\begin{eqnarray}
\Lambda_{stat}  &=& \overline{ {\rm Tr}\Big\{ \left(D^{\dagger} (t) - \overline{D^\dagger(t)} \right) \left(D(t) - \overline{D(t)} \right) \Big\} } \nonumber \\
&=&  \overline{ {\rm Tr}\Big\{D^{\dagger}D(t)\Big\} } - {\rm Tr} \Big\{\overline{D(t)}^2\Big\} . 
\label{eq:stat}
\end{eqnarray}
Starting from Eq. (\ref{eq:stocmfsimple}), 
the evolution of $\Lambda_{stat}$ over a small time step 
reads
\begin{eqnarray}
d \Lambda_{stat} &=& \frac{2dt}{\hbar} 
\Big\{ \overline{\left\langle  {Q}^2 \right\rangle_S} +  \overline{\left\langle  {B}^2 \right\rangle_E} \Big\},
\label{eq:statnomf}
\end{eqnarray}
where $\left\langle  {Q}^2 \right\rangle_S \equiv {\rm Tr_S (Q^2 \rho_S(t))}$ and $\left\langle  {B}^2 \right\rangle_E \equiv {\rm Tr_E (B^2 \rho_E(t))}$. Statistical errors associated with Eq. (\ref{eq:stocmfsimple}) have been estimated numerically and turn out to grow very fast in
time \cite{Zho05}. As a consequence, the stochastic process in the present form is useless to simulate physical situations and methods to reduce statistical errors should be used. 
  
To do so, it is worth to note that the stochastic equation of motion is not unique. Indeed, any stochastic process of the form: 
\begin{eqnarray}
\left\{
\begin{array} {lll}
d\rho_S &=& \frac{dt}{i\hbar}[H_S + Q \Delta_E , \rho_S] \\
 \\
&&+  
d\xi_S (Q - \Delta_S) \rho_S + d\lambda_S 
\rho_S (Q - \Delta_S)
\\
\\
d\rho_E &=& \frac{dt}{i\hbar}[H_E + B \Delta_S ,\rho_E]  \\
\\
&& +  
d\xi_E ({B} - \Delta_E ) \rho_E 
+ d\lambda_E \rho_E ({B} - \Delta_E ) 
\end{array}
\right.
\label{eq:stocmf},
\end{eqnarray}
where $\Delta_S(t)$ and $\Delta_E (t)$ are time-dependent parameters
leads to the same average evolution.} 
These stochastic equations also provide a reformulation of the initial system+environment problem. Indeed, we have 
\begin{eqnarray}
\overline{d\rho_S \otimes \rho_E} + \overline{ \rho_S \otimes d\rho_E} = \frac{dt}{i\hbar} [H_S + Q \Delta_E ,\rho_S \otimes \rho_E ] &&\nonumber \\
+ \frac{dt}{i\hbar} [H_E + B \Delta_S , \rho_S \otimes \rho_E ]  &&\nonumber \\
\overline{d\rho_S \otimes d\rho_E} = \frac{dt}{i\hbar} [ ({Q} -  \Delta_S )\otimes ({B} - \Delta_E ), \rho_S \otimes \rho_E ]. &&\nonumber
\end{eqnarray}  
Therefore, terms appearing in the deterministic part are exactly compensated by equivalent terms coming 
from the average over the noise. 
Accordingly, the evolution of the average density identifies with the exact equation of motion (\ref{eq:evoldtot}). 

Up to now, the flexibility has been essentially exploited by using
\cite{Lac05,Lac08b} 
\begin{eqnarray}
\Delta_E (t) =  \langle {B} (t) \rangle_E  , ~~~~ \Delta_S(t)  &=& \langle {Q} (t) \rangle_S.
\label{eq:newdelta}
\end{eqnarray}
This choice is justified by the fact that it directly appears when the Ehrenfest theorem is applied 
to separable total density for system or environment observables. 
By modifying the stochastic evolution, part of the coupling 
is already contained in the deterministic evolution. Accordingly, we do expect that the amount
of coupling to be treated by the noise is significantly reduced as well as the statistical errors. In the latter 
case, statistical fluctuations are given by:
\begin{eqnarray}
d \Lambda^{}_{stat} &=& \frac{2dt}{\hbar} 
\Big\{ \overline{(\langle  {Q}^2 \rangle_S- \overline{ \langle  {Q} \rangle}^2_S)} 
+ \overline{(\langle  {B}^2 \rangle_S- \overline{ \langle {B} \rangle}^2_S)}  \Big\} ,
\label{eq:statmf}
\end{eqnarray}
and are always smaller than the original ones (\ref{eq:statnomf}). 
As shown numerically in ref. \cite{Lac05},
introduction (\ref{eq:newdelta}) significantly reduces statistical fluctuations
and opens new perspectives for the application of the present framework. 

The modified stochastic theory has other advantages. For instance, 
the traces of densities are constant and remain equal to their initial values, i.e. $d{\rm Tr}(\rho_{S/E})= 0$. 
This greatly simplifies expectation values of system and/or environment observables. Indeed, denoting by $X$ a system operator, along 
a trajectory, we have 
\begin{eqnarray}
\left\langle  X \right\rangle &=& {\rm Tr_E} (X D(t)) = {\rm Tr}(\rho_E(t)) {\rm Tr_S} (X \rho_S(t)). 
\end{eqnarray} 
For stochastic processes with varying trace of densities, the 
observable evolution will contain terms coming from $d{\rm Tr}(\rho_{S/E})$ and cross terms coming from $d{\rm Tr}(\rho_{S/E}) 
d{\rm Tr} (X \rho_S(t))$. In the case considered here, we simply have 
\begin{eqnarray}
d\left\langle  X \right\rangle &=&  {\rm Tr_E}(\rho_E(t)) d {\rm Tr_S} (X \rho_S(t)). 
\end{eqnarray} 
{The QMC theory with centered noise overcomes the difficulty (ii) but does not help for (i) since the environment degrees of freedom should still 
be followed in time. In the next section, we show how irrelevant degrees of freedom can be projected out to obtain a closed 
stochastic master equation for the system only.  
}

\subsection{Reduced system density evolution and link with influence-functional theory}
\label{sec:master}
 
The stochastic formulation suffers a priori from the same difficulty as the total dynamics: the 
environment is in general rather complex and has a large number of degrees of freedom which can hardly 
be followed in time. In Eq. (\ref{eq:stocmf}), the influence of the environment 
on the system only enters through $\langle {B} (t) \rangle_E$. Therefore, instead of following the full environment 
density evolution, one can concentrate on this observable only. As shown in ref. \cite{Lac08}, the second equation in 
Eqs. (\ref{eq:stocmf}) can be integrated in time to give:
\begin{widetext}
\begin{eqnarray}
\left\langle {B}(t) \right\rangle_E &=& {\rm Tr_E}({B}^I(t -t_0) \rho_E(t_0))-
\frac{1}{\hbar}\int^t_0 D(t,s) \left\langle {Q} (s) \right\rangle_S ds 
-\int^t_0 D(t,s) du_E(s)  + \int^t_0 D_1(t,s) dv_E(s),
\label{eq:btime}
\end{eqnarray}
where $B^{I}$ denotes the operator $B$ written in the interaction picture while 
$D$ and $D_1$ are the memory function given by 
\begin{eqnarray}
D(t,s)  & \equiv & i \langle [ B(t), B(t-s)] \rangle_E,~~ {\rm  and} ~~
D_1(t,s) \equiv  \langle  \{ B(t)  , B(t-s)  \}_+ \rangle_E - 2 \langle  B(t) \rangle_E  \langle B(t-s) \rangle_E \label{eq:d1d2mf}.
\end{eqnarray}
A new set of stochastic variables  $dv_{S/E}$ and $du_{S/E}$ have been introduced through
$d\xi_{S/E} = dv_{S/E} - i du_{S/E}$ and $d\lambda_{S/E} = dv_{S/E} + i du_{S/E}$, and verify
\begin{eqnarray}
\overline{du_Sdu_E} &=& \overline{dv_Sdv_E} = \frac{dt}{2\hbar},  ~~~~
\overline{du_Sdv_E} = \overline{dv_Sdu_E} = 0. 
\label{eq:noiseusue}
\end{eqnarray}
Reporting the evolution of $\left\langle {B}(t) \right\rangle_E $ into the evolution of $\rho_S$, a closed 
stochastic equation of motion for the system density is obtained: 
\begin{eqnarray} 
d \rho_S = \frac{dt}{i\hbar} \left[ H_S , \rho_S \right] + dt [Q,\rho_S] \int^t_0 ds D(t-s) \left\langle  Q(s) \right\rangle_S 
+d \xi(t) [Q,\rho_S] + d\eta(t) \{Q - \left\langle  Q \right\rangle_S , \rho_S \} 
\label{eq:smfint}
\end{eqnarray}
with 
\begin{eqnarray}
d\xi(t) &=&  dt \int^t_0 D_1(t-s) dv_E(s) - dt \int^t_0 D(t-s) du_E(s) -idv_S(t) , ~~~d\eta(t) = du_S (t).
\end{eqnarray}
\end{widetext}
By integrating out the evolution of the environment, a new stochastic term is found that depends not only on the noise at time $t$ but also on its full history 
through the time integrals. Using second moments given by Eqs. (\ref{eq:noiseusue}) leads to:
\begin{eqnarray}
\overline{d\eta(t) d\eta(t')} &=& 0, 
\nonumber  \\
\overline{d\xi(t)d\eta(t') } &=& -\frac{dt}{2\hbar} \Theta (t-t')D(t-t'), \nonumber \\
\overline{d\xi(t)d\xi(t') } &=& -\frac{i dt}{2\hbar} D_1(|t-t'|), \nonumber
\end{eqnarray}
where $\Theta (t-t') = 1$ if $t>t'$ and $0$ elsewhere. Interestingly enough, the 
stochastic equation given by (\ref{eq:smfint}) identifies with the stochastic master 
equation obtained in ref. \cite{Sto02} using a completely different method based on the Feynman-Vernon 
influence functional theory \cite{Fey63}. It should however be kept in mind that different strategies to design 
the stochastic equation (see discussion in section \ref{sec:flex}) would have given a different 
stochastic master equation.

Despite the apparent complexity of Eq. (\ref{eq:smfint}), the QMC approach has been recently 
applied with success to the spin-boson model coupled to a heat bath of oscillators \cite{Lac08}. In particular, 
the introduction of (\ref{eq:newdelta}) seems to cure the numerical difficulties that have been encountered in this model \cite{Zho05}. 
By projecting the environment effect onto the system density evolution we do not need anymore 
to follow the environment density and we expect that a rather limited number of trajectories will be sufficient to accurately simulate 
the onset of dissipation and fluctuation in an open quantum system. Eq. (\ref{eq:smfint}) is the equation that is solved in practice.
It should be noted that the present stochastic process differs significantly from conventional approaches. Indeed, according to the 
noise properties, system densities are non-hermitian along a stochastic path. As a consequence, expectation value of observables 
are complex loosing their physical meaning before averaging over the stochastic path. Nevertheless, we illustrate in the following that the new theory can be a very powerful 
tool.

\section{Application}
\label{sec:appli}

The Caldeira-Leggett (CL) model \cite{Cal82} corresponds to a single harmonic oscillator 
coupled to an environment 
of harmonic oscillators initially at thermal equilibrium, i.e. 
\begin{eqnarray}
H_S = H_c + \frac{P^2}{2M} + \varepsilon \frac{1}{2}M\omega_0^2 Q^2, \\
H_E = \sum_{n}\left(\frac{p^2_n }{2m_n} + \frac{1}{2}m_n \omega^2_n x^2_n\right)
\end{eqnarray}
and $B \equiv -\sum_n \kappa_n x_n$ \cite{Bre02}. Here, $H_c = Q^2 \sum_n \frac{\kappa^2_n}{2m_n \omega^2_n}$ is the counter-term that 
insures that the physical frequency is $\omega_0$. In the following, $\varepsilon$ is either $+1$ (harmonic case) or $-1$ 
(inverted parabola case). Such a model can be solved exactly.

As shown in ref. \cite{Lac08}, the two functions $D$ and $D_1$ defined by Eqs. (\ref{eq:d1d2mf}) and estimated along the stochastic 
trajectories identify with the standard times correlation functions: 
\begin{eqnarray}
\hspace*{-0.2 cm}
D(\tau) &=&  2\hbar \int^{+\infty}_0 d\omega J(\omega) \sin(\omega \tau) , \label{eq:d1exp} \\
\hspace*{-0.2 cm} D_1(\tau) &=&  2\hbar \int^{+\infty}_0 d\omega J(\omega) \coth({\hbar \omega / 2k_B T}) \cos(\omega \tau) , \label{eq:d2exp}
\end{eqnarray}
where $\displaystyle J(\omega)\equiv \sum_n \frac{ \kappa^2_{n}} {2m_n \omega_n} \delta(\omega - \omega_n)$  
denotes the spectral density characteristic of the environment \cite{Leg87,Bre02}. In the following, a Drude spectral density \cite{Lac08}
\begin{eqnarray}
J(\omega) &=& \frac{2 M \eta}{\pi} \omega \frac{\Omega^2}{\omega^2+ \Omega^2},
\end{eqnarray}
 is considered where $M$ is the nucleon mass.

\subsection{Quantum Monte-Carlo method for parabolic potentials}
\label{sec:parab}
As a first illustration, the harmonic case ($\varepsilon = +1$) is considered. This case has been already studied in ref. \cite{Sto01} using the stochastic method proposed in ref. \cite{Sto02}.
In the CL model, starting from an initial Gaussian density, the system density remains gaussian along the
stochastic path. Therefore, the stochastic evolution of the system
density reduces to the first and second moments evolution of $\langle P\rangle$ and $\langle Q\rangle$ given by \cite{Lac08b}: 
\begin{eqnarray}
\left\{
\begin{array} {lll}
d\langle Q \rangle &=&\frac{\left\langle  P \right\rangle}{M }dt+ 2  du_S \sigma_{QQ}
\\
\\
d\left\langle P \right\rangle &=& - M\omega^2_0 \langle Q \rangle  dt 
- dt \left\langle  B \right\rangle 
+ 2 du_S \sigma_{PQ} 
-\hbar dv_S 
\\ 
\\
d\sigma_{QQ} &=& 2\frac{dt}{M} \sigma_{PQ} 
 \\
 \\
d\sigma_{PP} &=& 
-2 M\omega^2_0 dt \sigma_{PQ} \\
\\
d \sigma_{PQ} &=&  \frac{dt}{M} \sigma_{PP} - M\omega^2_0 \sigma_{QQ} dt  
\end{array}
\right. \label{eq:sme}
\end{eqnarray}
These equations illustrate the differences between the new exact reformulation and standard methods to treat dissipation. 
Generally, the noise enters into the evolution of $\langle P\rangle$ only and affects directly the second moment. Here, we see that second moments identify with the unperturbed ones while the random forces enter in both $\langle Q\rangle$ and $\langle P\rangle$. In addition, the noise is complex, which implies 
that observables make excursions into the complex plane. 
This stems from 
the specific noise used to design the exact formulation that leads to non-Hermitian densities along paths. 
Part of the conceptual difficulties in understanding the physical meaning of observable evolutions can be 
overcome by noting that if $\rho_S (t)$ 
belongs to the set of trajectories, by symmetry $\rho^\dagger_S (t)$ will
also belongs to the set. By grouping these two trajectories to estimate observables, real quantities are deduced.   

The exact evolution is obtained by averaging over different trajectories. For second moments, this leads to  
\begin{eqnarray}
\Sigma_{QQ} \equiv \overline{\langle Q^2 \rangle} - \overline{\langle Q \rangle}^2 =  \sigma_{QQ} + \overline{\langle Q \rangle^2} - \overline{\langle Q \rangle}^2 \label{eq:statq}
\end{eqnarray} 
where $\overline{\langle X \rangle}$ denotes the statistical average of quantum expectation values $\langle X\rangle$.
It is a particularity of the CL model that total fluctuation is recovered by simply adding up 
quantum and statistical fluctuations.

\begin{figure}[htb]
\begin{center}
\includegraphics[width = 7.cm]{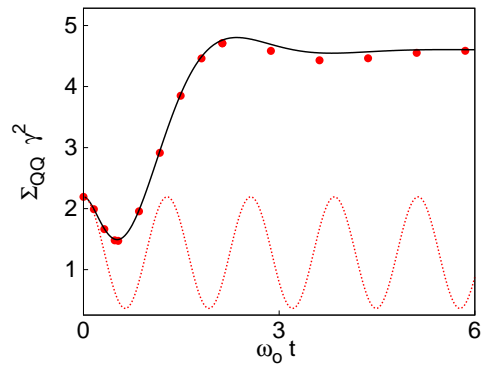}
\end{center}
\caption{(Color online) 
Evolution of $\Sigma_{QQ}$ (filled circles) obtained by averaging over $10^5$ trajectories.
This evolution is compared to the exact result (solid line) and to the quantum fluctuation $\sigma_{QQ}$ evolution (dotted line).
 Parameters of the simulation are $k_B T = \hbar \omega_0$, $\hbar \Omega = 5 \hbar\omega_0$, $\eta =0,5 \hbar \omega_0$ 
and $\hbar \omega_0 = 14 MeV$. The factor $\gamma$ defined as $\gamma^2 = \hbar/(2M\omega_0)$ equals here 
$\gamma = 1.216$ fm$^{-1}$.}
\label{fig:SMEsim2}
\end{figure}
An example of $\Sigma_{QQ}(t)$ evolution obtained using Eq. (\ref{eq:statq}) (red filled circles) 
is compared to the exact result (solid line) in figure \ref{fig:SMEsim2}. As an indication, 
the evolution of quantum fluctuation $\sigma_{QQ}$, which is identical for all trajectories, is also displayed (dotted line). 
Note that, Eq. (\ref{eq:smfint}) is already 
exact for the evolution of first moments $\langle P\rangle$ and $\langle Q\rangle$, even if the noise is omitted.
However, it completely fails to account for fluctuation.   
While the quantum evolution does not present any damping, the average evolution closely follows 
the exact solution. 
The harmonic oscillations in $\sigma_{QQ}$ are due to the fact that the width of the initial density 
differs from the width of the coherent state associated to the considered oscillator, i.e. $\sigma_{QQ}(0) \neq \hbar/(2M\omega_0)$. This is at variance with the simulation made in ref. \cite{Sto01}.     

\begin{figure}[htbp]
\begin{center}
\includegraphics[width = 0.5\textwidth]{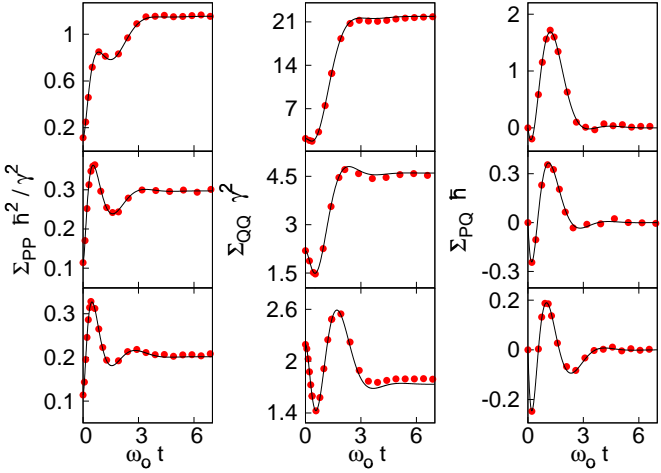}
\end{center}
\caption{(Color online) Evolution of $\Sigma_{PP}$ (left), $\Sigma_{QQ}$ (middle) and $\Sigma_{PQ}$ (right) obtained with $10^5$ trajectories are displayed with red filled circles as a function of time and systematically compared 
with the exact evolution (solid line). 
$k_B T= 5\hbar \omega_0$, $\hbar \omega_0$ and $0,1\hbar\omega_0$ are respectively shown 
from top to bottom. In all cases, $\eta =0,5 \hbar \omega_0$, $\hbar \Omega = 5 \hbar\omega_0$ and 
$\hbar \omega_0 = 14 MeV$. }
\label{fig:SMEsim}
\end{figure}

The accuracy of the quantum Monte-Carlo theory has been systematically investigated for various temperatures and coupling strengths.
In all cases, averaged evolutions could almost not be distinguished from the exact evolution. This is illustrated 
in figure \ref{fig:SMEsim} where $\Sigma_{QQ}$ (left), $\Sigma_{PP}$ (middle) and $\Sigma_{PQ}$ (right) 
are displayed as a function of time and compared to exact solutions for various temperatures. 
Figure \ref{fig:SMEsim}, clearly shows that the stochastic method properly includes all non-Markovian effects. In particular 
at low temperature, typically $k_B T < \hbar \omega_0$, and medium coupling constant $\eta$, large memory effect is 
expected. 

\subsection{Application of NZ and TCL}

Conjointly to the benchmark of quantum Monte-Carlo approaches, we also tested projection method either based on the Nakajima-Zwanzig \cite{Nak58,Zwa60,Bre02,Bre07a} or Time ConvolutionLess \cite{Shi77,Has77,Bre02,Bre07a} formalisms. Both theories provide  {\it a priori} 
exact 
re-formulations of the initial problem and lead to a closed master equation for the system density. However, they differ completely in the strategy and equation of motion used to incorporate memory effects. In the NZ case, the evolution of the system density at time $t$ depends 
on its full history (i.e. on $\rho_S(s)$ for all $s \leq t$). In the TCL case, the master equation is local in time and 
non-Markovian effects are treated in time-dependent transport coefficients. To illustrate the differences between our new QMC method and TCL, we remind below the corresponding local master equation, for $V = Q \otimes B$.
\begin{eqnarray}
 \hbar \frac{d}{dt}\rho_S(t) =\hspace{-1.5mm}&-&\hspace{-1.5mm} i[H_S,\rho_S(t)] - \frac{i}{2} \Delta(t) [Q , \{Q,\rho_S(t) \}]  \nonumber\\ \hspace{-1.5mm}&-&\hspace{-1.5mm} 2 i \lambda(t)[Q,\{P,\rho_S(t)\}] - \frac{D_{PP}(t)}{\hbar}[Q,[Q,\rho_S(t)]] \nonumber\\ \hspace{-1.5mm}& +&\hspace{-1.5mm} 2 \frac{D_{PQ}(t)}{\hbar}[Q,[P,\rho_S(t)]]. \label{eq:TR}
\end{eqnarray}
$\Delta(t)$, $\lambda(t)$, $D_{PQ}(t)$ and $D_{PP}(t)$ are time-dependent transport coefficients that contain memory effects. 
Similarly to the QMC case, the solution of the master Eq. (\ref{eq:TR}) is equivalent to follow first and second moments given by:
\begin{eqnarray}
\left\{
\begin{array} {llll}
\displaystyle
  \frac{d \langle Q \rangle}{dt} & =  &  
\displaystyle \frac{ \langle P \rangle  }{M} \\ \\
 
\displaystyle  \frac{d \langle P \rangle}{dt} & =  -&  
\displaystyle  M \omega_p^2(t) \ \langle Q \rangle - 2 \lambda(t) \langle P \rangle  \\ \\
\displaystyle  \frac{d\Sigma_{PP}}{dt} & = -& 

\displaystyle  2 M \omega_p^2(t)  \ \Sigma_{PQ} -4 \lambda(t) \Sigma_{PP}  
+ 2D_{PP}(t)   \\ \\
\displaystyle  \frac{d\Sigma_{QQ}}{dt} & = & 
\displaystyle 2 \frac{\Sigma_{PQ} }{M}          \\ \\
\displaystyle  \frac{d\Sigma_{PQ}}{dt} & = -&  
\displaystyle  M \omega_p^2(t) \ \Sigma_{QQ} - 2 \lambda(t) \Sigma_{PQ}  \\
 \ &  \hspace{4mm} + &  \displaystyle   \frac{\Sigma_{PP}}{M}+2D_{PQ}(t) 
\end{array} 
\right. , \nonumber 
\end{eqnarray}
with $\omega_p^2 (t)   =   \omega_0^2 + \Delta (t)$. 
In practice, the exact NZ or TCL theory cannot be exactly solved 
and an expansion in powers of the coupling constant is made. In the following, NZ2 (or TCL2) will refer to the expansion up to second order while NZ4 (or TCL4) will refer to the expansion is made up to fourth order. 
By neglecting higher orders in the coupling in NZ2 (resp. TCL2) or NZ4 (resp. TCL4), both theory are not exact anymore. In the following, the efficiency of each method is systematically discussed.

\begin{figure}[htb]
\begin{center}
\includegraphics[width = 0.5\textwidth]{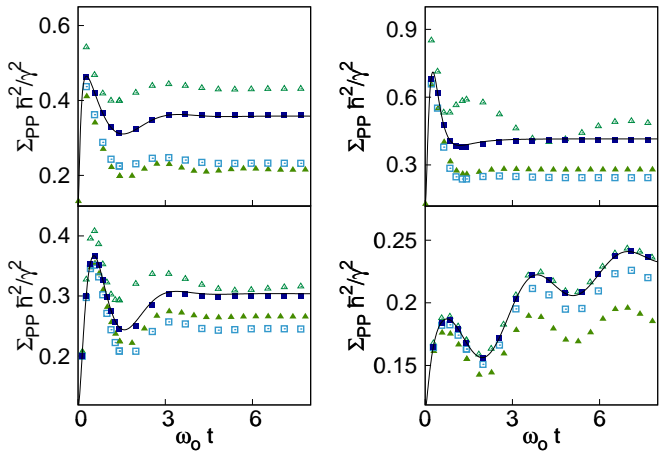}
\end{center}
\caption{(Color online) Evolution of $\Sigma_{PP}$ for different approximations: NZ2 (open triangles), NZ4 (filled triangles), TCL2 (open squares) and TCL4 (filled squares). The exact evolution is displayed with solid line.
In all cases, $\hbar \omega_0 = 14$ MeV, and $k_B T= \hbar \omega_0$ are used.  
The left side, corresponds to different cut-off frequencies: $\hbar \Omega = 20 \hbar \omega_0$ (top) and 
$\hbar \Omega = 5 \hbar \omega_0$ (bottom). In both cases, $\eta = 0,5 \hbar \omega_0$. In the right side, $\hbar \Omega = 10 \hbar \omega_0$
and different coupling strengths are used: $\eta =\hbar \omega_0$ (top) and  $\eta = 0.1\hbar \omega_0$ (bottom).}
\label{fig:tcl-nz}
\end{figure}

In Figure \ref{fig:tcl-nz}, the evolution of $\Sigma_{PP}$ for different cut-off frequencies $\hbar \Omega$ and coupling strengths 
$\eta$ are compared to the exact evolution (solid line). Explicit forms of the equation of motion in the NZ and TCL case can be respectively 
found in ref. \cite{Nak58,Zwa60,Bre02,Bre07a} and \cite{Shi77,Has77,Bre02,Bre07a}. 
Several important remarks could be drawn from this comparison: (i) In all cases, when 
the coupling strength is considered up to second order, NZ2 (open triangles) provides a better approximation than TCL2 (open squares). This might indeed be 
expected since NZ2 and TCL2 are respectively equivalent to the Born and Redfield master equation and the former contains a priori 
less approximations than the latter. (ii) While the TCL4 (filled squares) leads to a clear improvement compared to the TCL2, NZ4 (filled triangles) is in general worse than NZ2.
This is a known difficulty of NZ approach and was one of the motivation for the introduction of TCL method (see discussion in ref. 
\cite{Has77,Shi77,Bre02}). This stems from the fact that the order in perturbation in NZ cannot be
identified. For instance, NZ2 (resp. NZ4) contains orders in coupling constant greater than 2 (resp. 4). As a result, the NZ 
theory does not lead to better results when the "apparent" order in the coupling increases. The TCL method essentially cures this 
pathology and precise orders in the coupling can be selected. (iii) Rather large deviations between the exact 
and TCL2 are observed for different cut-off frequencies and coupling strengths. 
This issue is important since several theories have been recently developed along the line of TCL2 
to include memory effects in fusion and fission reactions \cite{Tak04,Ayi05,Was07}. 
{ Note that, the accuracy of TCL2 depends on different parameters used in the spectral density, in particular of the parameter $\hbar \Omega$. Here, we have used a value of the cut-off frequency between 10 and 20 MeV, which gives realistic dissipation and fluctuation for the fusion or fission mechanism\cite{Tak04,Ayi05,Sar08}.
}   
Our study clearly points out that a proper treatment of memory requires to include higher order 
effects.  
(iv) Finally, in all cases, TCL4 could not be distinguished from the exact result.  
As we will see, the efficiency of TCL4 is similar for the inverted parabola.

Since NZ method is not competitive, only the quantum Monte-Carlo and TCL methods are considered
in the following application. 

\subsection{Quantum Monte-Carlo method applied to inverted oscillators}

Several approaches have been recently developed to describe fusion and fission 
reactions \cite{Ari99,Boi03,Tak04,Ayi05,Was07,Sar08}. 
In these mechanisms, few collective 
degrees of freedom couple to a sea of internal excitations while passing an inverted  
barrier. At very low energy, both quantum and non-Markovian effects are expected to play a 
significant role. Most of the theory currently used start from quantum master equations 
deduced from TCL2. The quantum Monte-Carlo method offers a practical alternative which has 
similarities with path integrals theory. Path integrals are known to provide a possible framework 
to include dissipation while passing barriers (see for instance \cite{Cal81}). However, due 
to their complexity, only few applications have been made so far \cite{Fro90,Fro96}. 
We compare here the different approaches for inverted potential ($\varepsilon= - 1$).
  
\subsubsection{Initial conditions, trajectories and mean evolution}

Initially, we consider a Gaussian density with quantum width $\sigma_{QQ}(0) = 0.16$ fm$^2$ and $\sigma_{PQ} (0) = 0$ MeV.fm/c
and positioned on one side of the potential (here taken arbitrarily at $\langle Q(0)  \rangle= Q_0 > 0$ while 
the barrier height is is located 
at $0$ fm and is by convention taken as $V_B = 0$ MeV). The initial kinetic energy, denoted $E_K (0)$
is set by boosting the density with an initial momentum $ \langle P(0)  \rangle = P_0 < 0$. 

\begin{figure}[htb]
\begin{center}
\includegraphics[width = 0.5\textwidth]{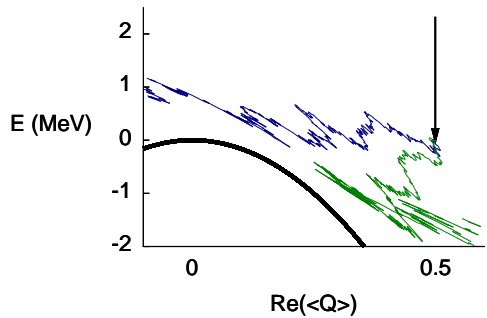}
\end{center}
\caption{(Color online) Evolution of $E(t)$ as a function of $Q(t)$ for two trajectories 
(dark and bright lines) with $\Delta E = 0$ MeV, $k_B T = 1 MeV$ and $\eta = 0.003MeV$. 
The black arrow indicates the initial position of the trajectories while the potential is also shown with  
bold line as a reference.}
\label{fig:traj}
\end{figure}

Contrary to the classical theory of Brownian motion, the notion of trajectories is
not so easy to tackle in the present Monte-Carlo framework. First, observables are complex. As   
mentioned in section \ref{sec:parab}, this difficulty can be overcomed by grouping trajectories by 
pairs which is equivalent to replace expectation of observables by their real parts. 
Second, it should be kept in mind that the present theory is a purely quantum theory where densities 
associated to wave-packets are evolved. Therefore, each trajectory should be interpreted in 
the statistical sense of quantum mechanics and contains many classical paths. Nevertheless, to 
visualize the trajectory we define the following energies: 
\begin{eqnarray}
E(t) = \frac{P(t)^2}{2m} - \frac{1}{2} m \omega^2_0  Q(t)^2
\end{eqnarray}
where $Q(t)$ and $P(t)$ denote the real part of  
 $\langle Q(t)  \rangle$ and $\langle P(t)  \rangle$ along the trajectory.
 An illustration of two trajectories, one passing the barrier and one reflected is shown in figure 
 \ref{fig:traj}. As illustrated in the following, it is convenient to group trajectories 
according to the quantity $\Delta E$
defined by 
\begin{eqnarray}
\Delta E &=& E(0) - V_B
\end{eqnarray}
which is nothing but the difference between total initial energy and barrier high. Both trajectories 
shown in figure  \ref{fig:traj} correspond to $\Delta E = 0$ MeV.

It is tempting to group trajectories into those passing the barrier and those reflected by the potential
to get information on the passing probability or passing time, however, it should be kept in mind that the 
present theory is fully quantal. Since each trajectories are associated with densities with quantum widths, both 
trajectories presented in Fig. \ref{fig:traj} contribute to the transmission probability.

\begin{figure}[htb]
\begin{center}
\includegraphics[width = 0.5\textwidth]{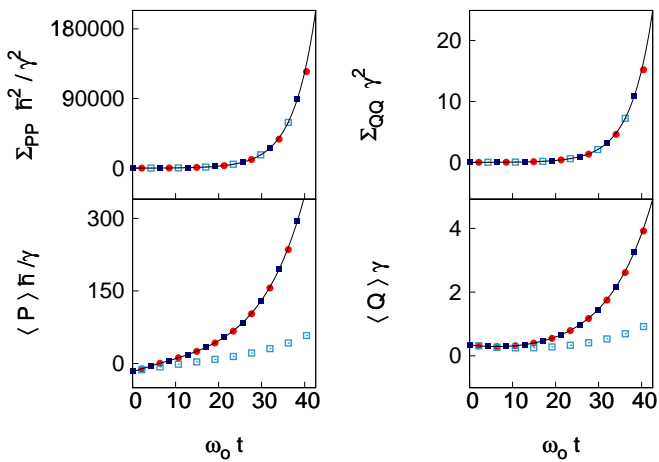}
\end{center}
\caption{(Color online) Evolution of $\langle Q\rangle$, $\langle P\rangle$, $\Sigma_{QQ}$ and $\Sigma_{PP}$ as a function of time obtained 
with quantum Monte-Carlo (filled circles), TCL2 (open squares) and TCL4 (filled squares). The exact evolution is also displayed 
with a solid line. Parameters of the simulations are $V_B = 4 MeV$, $\hbar\omega_0 =1 MeV$, $\eta = 0.03 MeV$, $\hbar \Omega =15 \hbar\omega$, $k_B T = 1 MeV$, $\Delta E = 0$ MeV and $\gamma = 0.402$ fm$^{-1}$. 
$10^5$ trajectories have been used for the quantum Monte-Carlo case.}
\label{fig:MedTHighC}
\end{figure}

The accuracy of different methods is illustrated in figure \ref{fig:MedTHighC} where evolutions of 
$\langle Q\rangle$, $\langle P\rangle$, $\Sigma_{QQ}$ and $\Sigma_{PP}$ are shown as a function of time.
Values of parameters retained for this figure are typical values generally taken in the nuclear context\cite{Sar07}. In all cases, including TCL2, second moments are well reproduced.
However, only TCL4 and the stochastic simulation provides a correct description of first moments. Calculations are shown here 
for $\Delta E = 0$ MeV. TCL2 provides a better and better approximation when $\Delta E$ increases while the disagreement 
increases below the barrier. This will be further illustrated below.

Different coupling parameters, cut-off frequencies and temperatures have been investigated showing 
that both TCL4 and quantum Monte-Carlo are very accurate theories leading always to results on top of the exact ones. It should 
be noted that the number of trajectories used in the stochastic approach to get small statistical errors 
is rather small (around $10^5$) which is quite encouraging for future applications.

\subsection{Transmission probability}

The accuracy of the method used to incorporate non-Markovian effects directly 
affects the predicting power of the theory. This aspect is illustrated here 
with the passing probabilities. Such a probability is a crucial ingredient 
in particular for models dedicated to the formation of very heavy elements 
\cite{Abe00,Bao02,Boi03,Tak04,Ayi05,Was07,Yil08}
and should be precisely estimated. 

\begin{figure}[htb]
\begin{center}
\includegraphics[width = 0.5\textwidth]{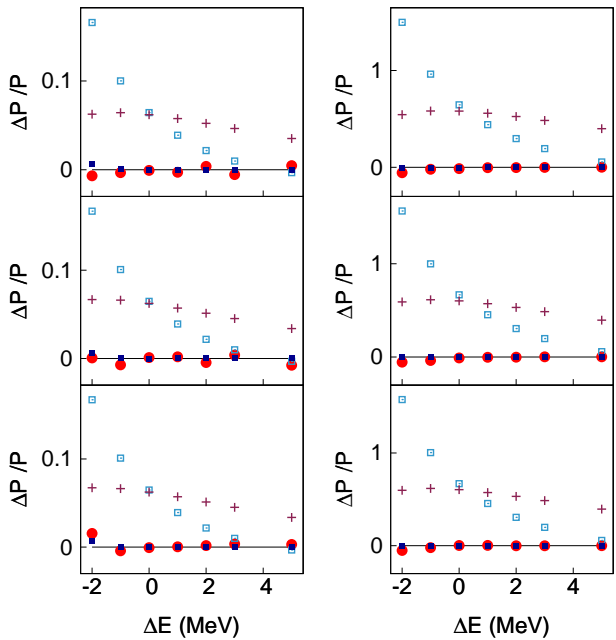}
\end{center}
\caption{(Color online) Evolution of $\Delta P / P$ as a function of $\Delta E$ calculated using 
quantum Monte-Carlo (filled circles), TCL2 (open squares), TCL4 (filled squares) and Markovian 
approximation (open crosses) for different coupling constant: $\eta = 0.003 MeV$ (left) and $\eta = 0.03 MeV$ (right). In both cases, 
 $T= 5\hbar \omega$, $\hbar \omega$ and $0,1\hbar\omega$ are shown from top to bottom.}
\label{fig:prob}
\end{figure}

The asymptotic passing probability is usually defined as:
\begin{eqnarray}
P(+\infty) &=&  \lim_{t\to + \infty}\frac{1}{2} {\rm erfc} \left( - \frac{|q(t)|}{\sqrt{2 \sigma_{qq} (t)}} \right). \label{eq:proba}
\end{eqnarray}
where $ q(t)$ and $\sigma_{qq} (t)$ denote the expectation value and second moment 
of $Q$ deduced from the considered theory. In the quantum Monte-Carlo case, these quantities identify with $\overline{\langle Q (t) \rangle }$ and $\Sigma_{QQ}(t)$ respectively. To quantify the precision of each theory, we have systematically investigated 
the difference between the estimated passing probability and the exact one using the parameter $\Delta P/P$:
\begin{eqnarray}
\frac{\Delta P}{P} &\equiv & \frac{P(+\infty) - P_{\rm ex}(+\infty)}{P_{\rm ex}(+\infty)}
\end{eqnarray}
where $P(+\infty)$ and $P_{ex}(+\infty)$ denote the results of the specific calculation considered and the exact one respectively.
Figure \ref{fig:prob} presents the evolution of $\Delta P/P$ as a function of $\Delta E$ obtained for different 
coupling strengths and temperatures for the quantum Monte-Carlo (filled circles), TCL2 (open squares), 
TCL4 (filled squares) cases. In this figure, the Markovian approximation is also shown by open crosses.
     
Once again, the TCL4 and the quantum Monte Carlo methods are in perfect agreement with the exact solution for any input parameters. 
Small differences sometimes observed between the stochastic approach and the exact value come from the limited number of trajectories 
used to obtain figure \ref{fig:prob}. Well above the barrier (here at least two times), TCL2 converges towards the exact case. However, at
low $\Delta E$, it turns out to be a rather poor approximation.
The difference seen in the TCL2 case can directly be traced back to the discrepancy already 
observed in figure \ref{fig:MedTHighC} and further confirms the difficulty of treating non-Markovian effects below the barrier. 
We can see that at lowest energy considered here, the error could be as large 
as $20\%$ in the weak coupling case and more than $100\%$ in the strong coupling limit. It is worth finally to mention 
that below the barrier, the Markovian limit gives an even better result that the TCL2 case.  

\section{Summary and discussion}

In this work, the quantum Monte-Carlo approach recently proposed in ref. \cite{Lac08} to incorporate exactly non-Markovian effects is introduce and  
applied to the case of harmonic potentials coupled to a heat-bath. 

For both non-inverted and inverted potentials, the new technique is rather effective to reproduce the exact evolution 
with a rather limited numbers of trajectories. Other methods have also been benchmarked.  
The TCL2 method, which is now widely used in nuclear physics to estimate 
passing probabilities, turns out to deviate 
significantly from the expected result especially below the barrier and
even in the weak coupling regime. To properly treat the dynamics of barrier transmission, higher
orders in the coupling strength should be incorporated. TCL4 gives very good agreement with the exact 
evolution in all cases considered here. The conclusion of our present work is that both 
quantum Monte-Carlo approach and TCL4 could be considered as good candidates to include memory 
effects for situations of interest in nuclear physics. Henceforth, the TCL2 method 
which is widely used nowadays should be replaced by TCL4.
The possibility to use  stochastic formulation that are exact in average 
open new perspectives to describe 
a system coupled to a complex environment. The application to harmonic potential 
gives interesting insight into such a theory. Application to more general potential 
turns out to be less straightforward with the appearance of spikes which 
have been already observed in several formalism where non linear stochastic 
equations appear \cite{Gar00}. To make these theories more versatile, new methods 
like the one proposed recently in ref. \cite{Koc08} could be used.\\

\begin{acknowledgments}
We thanks D.Boilley for his careful reading of the manuscript and G.Adamian and N.Antonenko for fruitful discutions.
\end{acknowledgments}

\bibliography{biblio_persotot}

\end{document}